\input{aipcheck}

\documentclass[
    ,final            
  ]
  {aipproc}

\layoutstyle{8x11double}
\usepackage{amsmath}

\begin{document}

\title{Superconducting Junctions with Ferromagnetic,
Antiferromagnetic or Charge-Density-Wave Interlayers}

\author{Yuri Barash}{
address={Institute of Solid State Physics, Russian Academy of Sciences,
Chernogolovka, Moscow reg., 142432 Russia}
}

\author{I. V. Bobkova}{
address={Institute of Solid State Physics, Russian Academy of Sciences,
Chernogolovka, Moscow reg., 142432 Russia}
}

\author{Brian M. Andersen}{
address={Department of Physics, University of Florida,
Gainesville, FL 32611 USA} }

\author{T. Kopp}{
address={Center for Electronic Correlations and Magnetism, University of
Augsburg, D-86135 Augsburg, Germany}
}

\author{P. J. Hirschfeld}{
address={Department of Physics, University of Florida,
Gainesville, FL 32611 USA} }

\begin{abstract}
Spectra and spin structures of Andreev interface states and the
Josephson current are investigated theoretically in junctions
between clean superconductors (SC) with ordered interlayers. The
Josephson current through the ferromagnet-insulator-ferromagnet
interlayer can exhibit a nonmonotonic dependence on the
misorientation angle. The characteristic behavior takes place if
the $\pi$ state is the equilibrium state of the junction in the
particular case of parallel magnetizations. We find a novel
channel of quasiparticle reflection (Q reflection) from the
simplest two-sublattice antiferromagnet (AF) on a bipartite
lattice. As a combined effect of Andreev and Q reflections,
Andreev states arise at the AF/SC interface. When the Q reflection
dominates the specular one, Andreev bound states have almost zero
energy on AF/ s-wave SC interfaces, whereas they lie near the edge
of the continuous spectrum for AF/d-wave SC boundaries. For an
s-wave SC/AF/s-wave SC junction, the bound states are found to
split and carry the supercurrent. Our analytical results are based
on a novel quasiclassical approach, which applies to interfaces
involving itinerant antiferromagnets. Similar effects can
take place on interfaces of superconductors with charge density
wave materials (CDW), including the possible d-density wave state
(DDW) of the cuprates.
\end{abstract}

\maketitle


Superconducting heterostructures involving ferro- and/or
antiferromagnets manifest unusual properties associated with spin
effects, and are of both fundamental interest and important for
technological applications. Superconductor-ferromagnet-superconductor
(SC/F/SC) junctions are known to display $0-\pi$
transitions with varying the temperature or the interlayer width.
We have demonstrated theoretically that the $0-\pi$ transition can
show up also at fixed temperature and interlayer width in
superconducting junctions with a three-layer FIF interface, as a
function of the misorientation angle between the magnetizations of two
F layers separated by the insulating barrier \cite{bbk}. The
dependence of the Josephson current on the misorientation angle
$\varphi$ becomes especially simple in the tunneling limit, when
the critical current takes the form
$J_c(T,\varphi)=J_c^{(p)}(T)\cos^2(\varphi/2)+J_c^{(a)}(T)
\sin^2(\varphi/2)$.\,Here $J^{(p)}(T) \& J^{(a)}(T)$ are critical
currents in tunnel junctions with parallel and antiparallel
orientations of the exchange fields in the three-layer interface.
The $0-\pi$ transition can take place with varying $\varphi$, if
$J^{(p)}_{c}(T)$ and $J^{(a)}_{c}(T)$ have opposite signs. This is
the case when the junction with parallel magnetizations is in the
$\pi$-state, since always $J^{(a)}_{c}(T)>0$. The transition
results in a nonmonotonic dependence of
$|J(T,\varphi)|$ on $\varphi$. This effect can be used for
switching the junction from the zero state to the $\pi$ state by
varying the misorientation angle. The angle is
changed in the FIF interlayer with applied magnetic field, if
it is larger than the coercive force in one of the F layers and
less than in the other layer.

Many fundamental and practical problems involve interfaces with
antiferromagnets. In particular, many of the properties of HTSC
cuprate materials probably arise from a competition between
antiferromagnetic and superconducting order, and many situations
involve such natural or fabricated boundaries. We have studied
interfaces between itinerant antiferromagnets and
normal metals or superconductors and demostrated that a new
spin-dependent channel of quasiparticle reflection, the so-called
$Q$ reflection, occurs on the interfaces \cite{bobkova05}.
Parallel to the interface, the momentum component of low-energy
normal-metal quasiparticles changes by ${Q}_{y}$ in a $Q$
reflection event, where $\mathbf{Q}$ is the wave-vector of the
antiferromagnetic pattern and $y$ is the direction parallel to the
interface. Assuming comparatively small Fermi velocity mismatches
and taking into account the nesting condition
$E_F(\mathbf{p+Q})=-E_F(\mathbf{p})$ in itinerant
antiferromagnets, one can see within the mean-field tight-binding
model on the half-filled square lattice that normal metal
quasiparticles with energies less than or comparable to the
antiferromagnetic gap change their momenta by $\bf Q$ and reverse
the signs of their velocities in a $Q$ reflection event. Consequently,
such quasiparticles experience spin-dependent retroreflection at
antiferromagnet-normal metal (AF/N) transparent interfaces.

Quasiparticle bound states below the AF and SC gaps at AF/SC
interfaces arise as a combined effect of Andreev and $Q$
reflections. Among a variety of subgap states, low-energy states
$E_B\ll {\rm min}\{m,\Delta\}$ are of special interest since they
can result in low-temperature anomalies in the Josephson critical
current, as well as low-bias anomalies in the conductance. Here
$m$ and $\Delta$ are the sublattice electronic magnetization and
the superconducting $s$-wave or $d$-wave order parameters. In the
absence of the interface potential $h$, the dispersive bound state
energies at the $(110)$ and $(100)$ AF - $s$-wave superconductor
(AF/sSC) interfaces can be represented as
$E_b(\mathbf{k}_F)=\pm\Delta_s\sqrt{R_{sp}( \mathbf{k}_F)}$. Here
$R_{sp}(\mathbf{k}_F)$ is the normal state reflection coefficient
for specular reflection from the interface, which occurs even in
the absence of any interface potentials due to a mismatch of Fermi
velocities in the AF and the sSC. Since normal-metal states are
presumably identical in the left and right halfspaces, under the
conditions $\Delta\ll m,\,t$, the mismatch in the model controls the
parameter $m/t$. If the magnetic order parameter $m$ is much less than
the hopping matrix element $t$, $Q$ reflection dominates
$R_{sp}(\mathbf{k}_F)\ll 1$, and bound state energies almost
coincide with the Fermi level $|E_b|\ll\Delta_s\le m$. In
particular, for the (110) interface on the square lattice we find
${R_{sp}(\mathbf{k}_F)=
\left[1+\left({\sqrt{2}v_{F,\perp}(\mathbf{k}_F)}/{am}\right)^2\right
]^{-1}}$, where $a$ is the lattice spacing and
$v_{F,\perp}(\mathbf{k}_F)$ the normal-state Fermi velocity
component along the boundary normal.
\begin{figure}[!tbh]
\includegraphics[width=6.5cm]{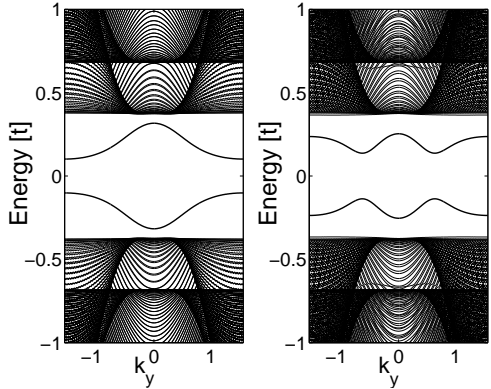}
\caption{Eigenvalues for the (100) AF/sSC interface as a function
of $k_ya$ ($a$ is the lattice constant) for (a) $\mu=0.0$ and $h=0.0$
and (b) $h=2.0t$. Order parameter values in the bulk:
$\Delta_{s,b}=0.4t$, $m_b=0.7t$. Here, one sees explicitly the
presence and dispersion of the bound state band inside the
gap.\label{bandsAFssc100}}
\end{figure}
In Fig.~\ref{bandsAFssc100} we plot the quasiparticle spectrum as
obtained from the selfconsistent eigenvalues of the Bogoliubov-de
Gennes equations. Interface bound states show up inside the main gap
of the spectrum as a distinct band, which disperses with the momentum
component $k_y$ along the interface. The two gap edges seen in Fig.
\ref{bandsAFssc100} are associated with the superconducting (lesser)
and the antiferromagnetic (larger) gaps. Interface potentials $h$
present near the interface tend to suppress the bound states
resulting from Q-reflection and move their positions towards the gap
edge. In the regime where $h$ is of the order of $t$, we find that
the main effect of the specular reflection channel is to cause a
stronger dispersion of the bound state energy. One can identify
additional extrema in the wave vector dependence of the bound state
energy $E(k_y)$. A typical example is seen in Fig. \ref{bandsAFssc100}
where $h=2.0t$. The new stationary points in the dispersion lead to
additional LDOS peaks near the interface.

Dispersive bound state energies on an antiferromagnetic - $d$-wave
superconductor interface (AF/dSC) can be represented as $E_b(
\mathbf{k}_F)=\pm\Delta_d(\mathbf{k}_F)\sqrt{R_{Q}(\mathbf{k}_F)}$.
They lie near the edges of the continuous spectrum, when $Q$ reflection
dominates. This contrasts with the case of a $(110)$ surface of a
dSC confined with an impenetrable wall, where zero-energy
Andreev states are formed.

For an sSC/AF/sSC junction, due to a finite width $l$ of the AF interlayer,
the low-energy bound states are split and carry the
supercurrent. If no potential barriers are present on the boundaries,
and $l\ll\xi_s$, $\Delta_s\ll m \ll t$, we find the following
energies for interface states: $\varepsilon_B=\pm\sqrt{D}|\Delta_s
\cos(\chi/2)|$, where $\chi$ is the order parameter phase difference,
$D(k_y)=4K(k_y)(K(k_y)+1)^{-2}$ is the transparency of the N/AF/N
junction and $K(k_y)=\exp(2ml/|v_{F,x}(k_y)|)$. For large interlayer
width, $K,D\ll 1$, there are low-energy states in the junction which
result in low-temperature anomalous behavior of the critical current.

Similar effects for CDW/SC interfaces have been studied recently in
\cite{bobkova205}. Subgap Andreev states arise at CDW/dSC and DDW/sSC
interfaces. At the same time there are no subgap states at CDW/sSC
interfaces due to the absence of interface-induced pair-breaking
processes. The interface states also do not arise at DDW/dSC interfaces
since pair-breaking effects from DDW and dSC compensate each other in
this case. In dSC/CDW/dSC and sSC/DDW/sSC Josephson junctions, the interface
low-energy bound states are split and strongly influence the Josephson
current.


This work was supported by grant
NSF-INT-0340536 (I.V.B., P.J.H., and Yu.S.B.), and by ONR grant
N00014-04-0060 (P.J.H and B.M.A). I.V.B. and Yu.S.B. also acknowledge
the support by grant RFBR 05-02-17175. T.K. thanks the support by the
DFG through SFB 484, DAAD D/03/36760 and BMBF 13N6918A.


\bibliographystyle{aipproc}   

\bibliography{sample}


\end{document}